\documentclass[conference]{IEEEtran}
\IEEEoverridecommandlockouts
\usepackage{cite}
\usepackage{amsmath,amssymb,amsfonts}
\usepackage{algorithm}
\usepackage{algorithmic}
\usepackage{graphicx}
\usepackage{textcomp}
\usepackage{xcolor}
\usepackage{float}
\usepackage{svg}
\usepackage{soul}
\usepackage{booktabs}
\usepackage{todonotes}
\usepackage[citecolor=blue,colorlinks=true,linkcolor=blue]{hyperref}%
\usepackage{array}

\usepackage{booktabs}
\usepackage{threeparttable}
\usepackage{makecell}  
\usepackage{amsmath}
\usepackage[dvipsnames]{xcolor}

\def\BibTeX{{\rm B\kern-.05em{\sc i\kern-.025em b}\kern-.08em
    T\kern-.1667em\lower.7ex\hbox{E}\kern-.125emX}}
\renewcommand\baselinestretch{0.970}
\begin{document}

\title{RQP: Resource-Oriented Quantiser Pruning for Neural Networks on FPGAs
\thanks{This work was supported by the Sustainable Energy Authority of Ireland under Grant number 24/RDD/1170.
}}

\author{
Changhong Li,
Biswajit Basu,
Shreejith Shanker \\ 
Reconfigurable Computing Systems Lab, Electronic \& Electrical Engineering\\
Trinity College Dublin, Ireland\\
Email: \{lic9, basub, shreejith.shanker\}@tcd.ie
\vspace{-3mm}
}

\maketitle

\begin{abstract}
High granularity quantisation (HGQ) exploits weight‑level quantisation and pruning to design resource‑efficient neural network accelerators, achieving an attractive trade‑off between accuracy and hardware utilisation.
HGQ is particularly well-suited to FPGA‑based edge neural network applications.
Standard HGQ workflow starts from a high‑precision model and progressively reduces bit‑width, guided by gradient‑based optimisation to outline the Pareto frontier.
This monotonic, irreversible pruning process is computationally intensive and can overlook the optimal subnetwork for a given resource level.
We propose a resource-oriented one-shot quantiser pruning that brings the network directly close to the target search space, and then use bidirectional $\beta$ scheduling for fine-tuning to enable a more refined scan of the Pareto frontier.
Validated on the jet substructure classification (JSC) task, our method reduces the search cost by up to $20.58\times$ compared with monotonic resource reduction in standard HGQ workflows, while achieving a competitive Pareto frontier and final network configuration.

\end{abstract}

\begin{IEEEkeywords}
Accelerator, Unstructured Sparsity, Model Compression, Machine Learning, Field Programmable Gate Arrays, Quantised Neural Nets
\end{IEEEkeywords}

\section{Introduction}\label{introduction}
The growing convergence of deep learning and edge computing has made low-latency neural network inference on specialised hardware increasingly important~\cite{singh2023edge}.
Many scientific computing and fast control applications demand microsecond or even sub-microsecond latency, with tight throughput and on-chip resource constraints.
Typical examples include high-energy physics experiments~\cite{duarte2018fast}, quantum circuit control~\cite{di2025end}, and high-frequency trading~\cite{kao2022fpga}.
FPGAs are an attractive platform for ultra-low-latency neural network acceleration, benefiting from their flexibility and massive parallelism.

\begin{figure}[htbp]
    \centering
    \includegraphics[width=0.48\textwidth]{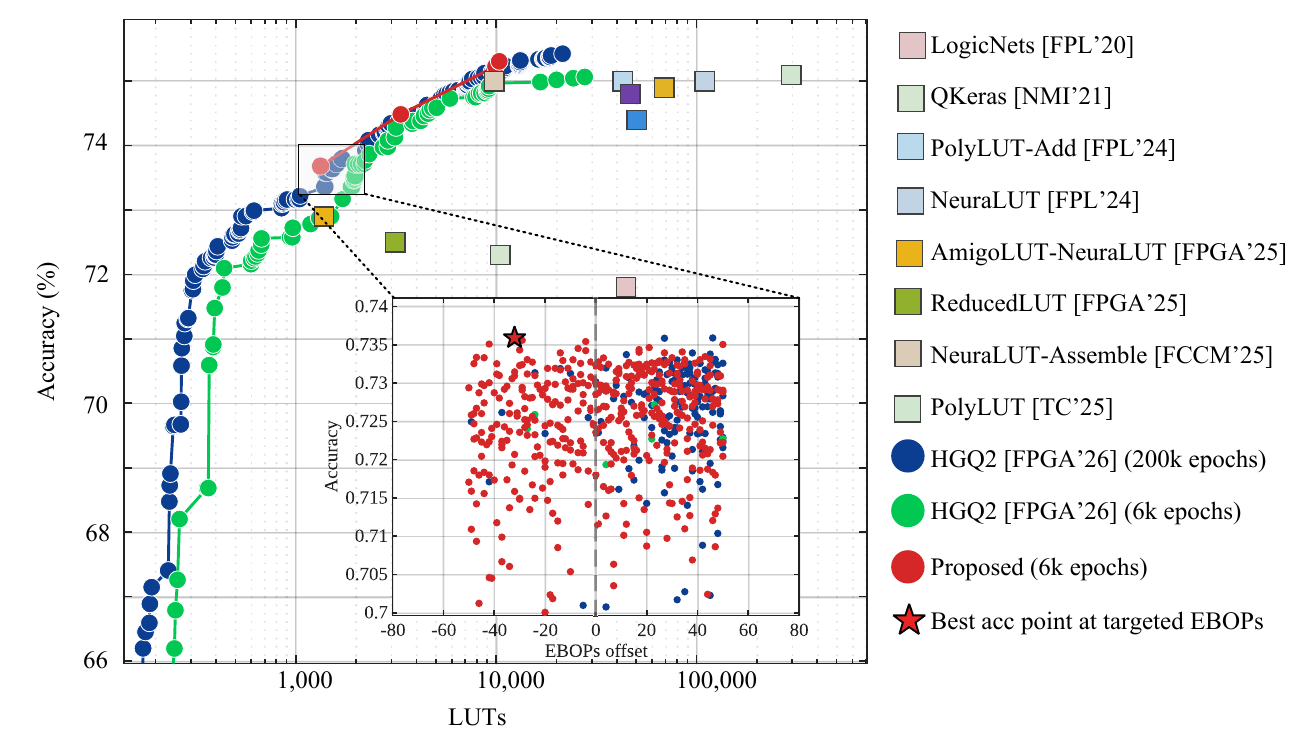}  
    \caption{Accuracy versus LUT consumption for High Level Features (HLF) JSC models on the CERNBox dataset. Square markers correspond to prior implementations. Green and blue circles denote the frontiers identified by HGQ through monotonic resource reduction with 200k and 6k training epochs. Red circles denote the frontiers achieved by RQP with 6k epochs.}
    \label{fig:mainpareto} \vspace{-5mm}
\end{figure}

High-accuracy neural networks often rely on deeper architectures and greater model complexity, whereas tight latency constraints require highly parallel hardware implementations.
These competing requirements make it difficult to deliver both strong predictive accuracy and strict real-time behaviour within limited on-chip resources.
The trade-off shown in Fig.~\ref{fig:mainpareto} for an example scenario (using the CERNBox dataset), capturing the Pareto frontiers of accuracy and resource usage achievable using existing techniques and the proposed Resource-Oriented Quantiser Pruning (RQP).  
%
Model compression and hardware software co-design have become essential to address it.

LUT-based networks, such as LogicNets~\cite{umuroglu2020logicnets}, LUTNet~\cite{wang2019lutnet}, PolyLUT~\cite{andronic2023polylut}, NeuraLUT~\cite{andronic2024neuralut, andronic2025neuralut} and AmigoLUT~\cite{weng2025greater}, map neurons directly to logical lookup table operations to exploit FPGA primitives.
On small-scale models, these methods can achieve low latency with efficient resources.
However, the scalability of such methods has not been fully verified.
Gradient-based optimisation becomes more difficult as the network grows, while hardware cost rises rapidly with the input degree of each neuron.
More conventional approaches rely on quantisation-aware training and pruning in standard neural networks, reducing numerical precision and effective computation to lower computational and storage costs~\cite{hawks2021ps}.
However, quantisation and pruning are often performed only at the layer or channel level because control overhead and memory access constraints limit finer-grained implementations.
These structural patterns narrow the compression space and cause avoidable accuracy loss.


\begin{figure*}[t]
    \centering
    \includegraphics[width=0.98\textwidth]{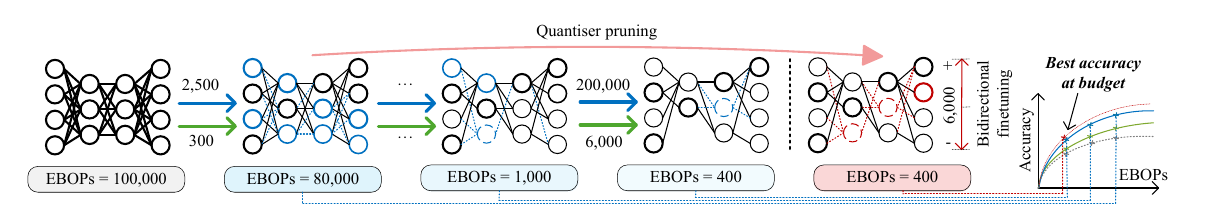}  
    \caption{RQP overview: The blue and green trajectories illustrate the progressive reduction of weight bit width, activation bit width, and the corresponding EBOPs during HGQ training, together with the associated pruning process. In contrast, the red trajectory denotes RQP, which starts from a high bit-width network, applies one-shot continuous quantiser bit-width pruning to reach the vicinity of the target budget directly, and then performs bidirectional resource adjustment to search for the optimal subnetwork near the target resource budget.}
    \label{fig:mainflow} \vspace{-4mm}
\end{figure*}

HGQ~\cite{sun2026hgq} addresses this limitation by enabling gradient-based optimisation of per-weight bit widths.
HGQ can optimise task accuracy and on-chip resource usage at the same time with differentiable resource proxies, allowing the training process to search for favourable accuracy-resource trade-offs directly.
However, existing HGQ training methods, which are similar to progressive pruning, still suffer from efficiency and search-space issues.
In conventional HGQ pipelines, training typically starts from a high-precision network and progressively steers the model toward lower resource configurations under increasing resource pressure during optimisation until the target budget is reached.
Much of the search efforts are wasted on intermediate resource regions.
The contraction schedule and training duration must be tuned carefully, as excessive resource pressure can cause the network to collapse prematurely.
This monotonic compression process also faces path dependence.
Once connections are pushed to low bit-widths or removed in the early training stage, their gradient paths are then restricted, and may miss better subnetworks.

To address this gap, we propose RQP, a resource-oriented one-shot continuous quantiser pruning method. 
Unlike conventional one-shot binary mask pruning~\cite{lecun1989optimal, lee2018snip, tanaka2020pruning, wang2020picking, lucas2024preserving}, the proposed method acts directly on per-weight quantiser bit widths in a continuous manner, projecting a high-precision model into the vicinity of the search space associated with the target resource budget.
This avoids redundant intermediate compression stages and allows finer-grained optimisation around the desired deployable point.
We also introduce a bidirectional $\beta$ scheduling strategy to mitigate path dependence and the locking effect, thereby enabling the exploration of better subnetworks.

The main contributions of this paper are as follows:
\begin{itemize}
    \item One-shot quantiser pruning with spectral constraints to obtain trainable quantised sparse subnets near the target resource budget to reduce search cost.
    
    \item An annealing-inspired bidirectional $\beta$ scheduler that alleviates compression locking and enables denser exploration around the target budget.
    
    \item Validated in a multi-budget accelerator training setting for the JSC classification task, RQP improves the training efficiency of reaching the target resource budget by up to $20.58\times$ over monotonic resource reduction, while providing a denser frontier space and competitive accuracy.
\end{itemize}

The remainder of this paper is organised as follows.
Section~\ref{methodology} describes our pruning and fine-tuning algorithms.
Section~\ref{results} discusses the experimental results from case studies on the model training and accelerator generation. Finally, Section~\ref{conclusion} concludes the paper.
\section{Methodology and Design} \label{methodology}
HGQ~\cite{sun2026hgq} unifies the quantisation and pruning process by optimising bit-widths during training, where zero bit-widths are naturally driven toward pruning, thereby bringing the granularity of quantisation and pruning to the bit-width level.
The resource proxy defined in Eq.~\ref{eq:ebops} estimates hardware cost derived from bit widths as Effective Bit Operations (EBOPs).
$\mathcal{M}=\{(i,j)\}_{n}$ denotes the set of all multiplication operations between operands with bit-widths $b_i$ and $b_j$ in the model, and $\mathcal{A}=\{(k,l)\}_{m}$ denotes the set of all explicit addition, subtraction, or muxing operations between operands with bit-widths $b_k$ and $b_l$.
Following the empirical validation in~\cite{sun2026hgq}, LUT utilisation can be estimated as
$\mathrm{LUT} \approx \exp\left(0.985 \log(\mathrm{EBOPs})\right)$, which enables resource control during training.

\begin{equation}
\mathrm{EBOPs} =
\sum_{i,j \in \mathcal{M}} b_i b_j
+
\sum_{k,l \in \mathcal{A}} \max(b_k, b_l)
\label{eq:ebops}
\end{equation}

The loss function is defined as in Eq.~\ref{eq:loss}. In addition to the task loss, the $\gamma$ regularisation on bit-widths is introduced to prevent unbounded bit-width growth.
Model size is primarily controlled by $\beta$.

\begin{equation}
L = L_{\mathrm{base}} + \beta \cdot \mathrm{EBOPs} + \gamma \cdot \sum \text{bit-widths}
\label{eq:loss}
\end{equation}

\begin{figure}[htbp]
    \centering
    \includegraphics[width=0.48\textwidth]{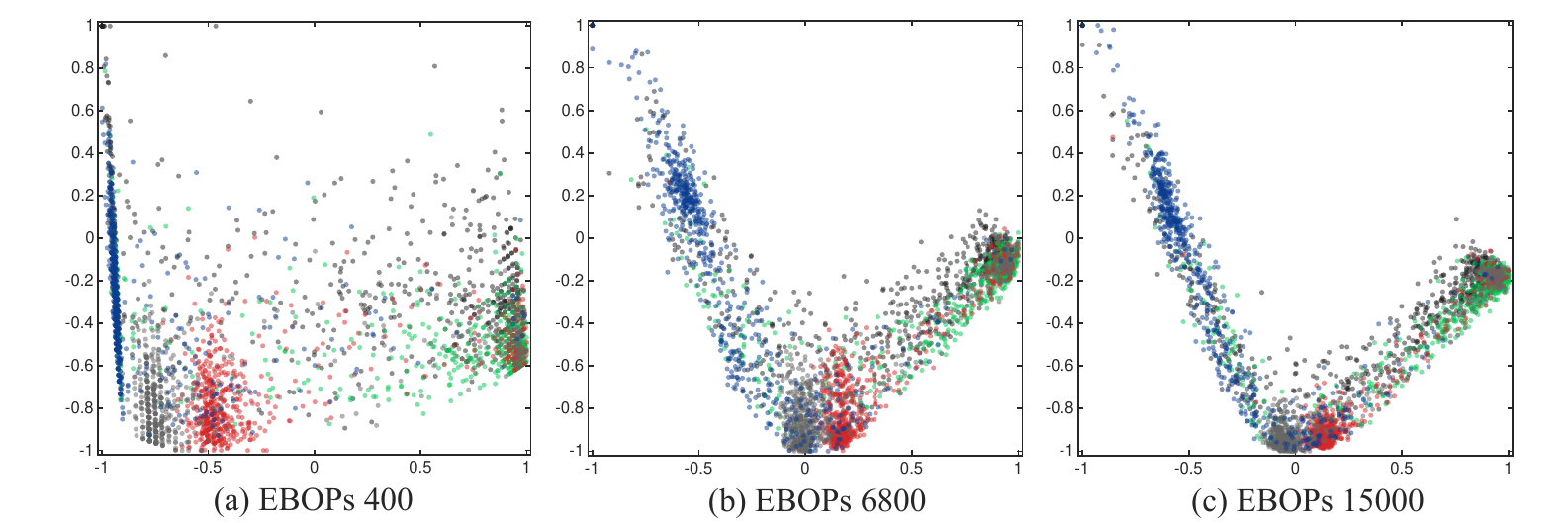}  
    \caption{2D PCA visualisation of the first-layer forward activation features for models of different sizes. At high resource budgets, the early layers primarily serve as feature extractors, while at low resource budgets, part of the classification function moves to earlier layers.}
    \label{fig:feature}
\end{figure}

In practice, progressively increasing the $\beta$ pressure drives the model smaller and thereby the Pareto frontier explored. 
Intermediate models along the optimisation path consume substantial training resources, and their layer functions and feature representations evolve progressively, as illustrated in Fig.~\ref{fig:feature}. However, these path models, which are hard to bypass, do not directly improve the performance of the final deployed model, and the path dependence of this long monotonic process may cause the optimisation to miss better subnetworks.

RQP adopts the resource proxy and the co-optimisation loss function of HGQ as its foundation, but replaces the progressive training with a one-shot quantiser pruning and bidirectional $\beta$ scheduler, as illustrated in Fig.~\ref{fig:mainflow}.
One-shot quantiser pruning extracts the subnetwork from a high-budget model. 
It operates directly on continuous quantiser bit widths instead of conventional discrete weight sparsity masks, thereby enabling finer-grained pruning and explicitly optimising the model’s implementation resource.
The pruned model is then fine-tuned by the bidirectional $\beta$ scheduler near the target budget. This optimisation can escape locally optimal topologies, perform a finer-grained scan near the target resource region, and reach a better deployable frontier.

\subsection{One-shot quantiser pruning with spectral constraints}
RQP employs one-shot pruning to project a pre-trained high-resource HGQ model into the target resource region. 
This pruning procedure consists of two consecutive steps: continuous bit-width reallocation to reshape the layer-wise resource distribution, and spectral-constrained topology compensation to improve the subnet’s trainability under the target budget.

Given a pretrained quantised model $\mathcal{N}$ and a target EBOPs budget $E^\star$, Algorithm~\ref{alg:target_bw_reallocation} describes the target-budget bit-width reallocation.
Let $c_{l,ij}$ denote the EBOPs cost per bit width at weight $(i,j)$ of layer $l$, and let $\mathbf{B}_l^{\mathrm{pre}}$ denote the pretrained bit-width matrix. 
The current budget is approximately evaluated as $E_{\mathrm{cur}}=\sum_l\sum_{i,j} c_{l,ij} B_{l,ij}^{\mathrm{pre}}$, from which a global contraction ratio $\alpha$ is obtained. 
For each prunable layer, the mean pretrained bit-width over active weights is used to define a layer statistic and derive a layer-wise scaling factor.
The pretrained bit-width matrix of each layer is rescaled accordingly within the valid range, yielding the reallocated bit-width matrix $\mathbf{B}_l^{(0)}$.

\begin{algorithm}[h]
\caption{Target-budget bit-width reallocation}
\label{alg:target_bw_reallocation}
\begin{algorithmic}[1]
\STATE \textbf{Input:} pretrained quantised model \(\mathcal{N}\), target budget \(E^\star\)
\STATE \textbf{Output:} reallocated bit widths $\{\mathbf{B}_l^{(0)}\}$
\STATE $E_{\mathrm{cur}} \leftarrow \sum_l \sum_{i,j} c_{l,ij} B_{l,ij}^{\mathrm{pre}}$
\STATE $\alpha \leftarrow E^\star / E_{\mathrm{cur}}$
\FOR{each prunable layer $l$}
    \STATE $\Omega_l \leftarrow \{(i,j)\mid B_{l,ij}^{\mathrm{pre}} > 0\}$
    \STATE $r_l \leftarrow \operatorname{mean}\{B_{l,ij}^{\mathrm{pre}} \mid (i,j)\in\Omega_l\}$
\ENDFOR
\STATE $\bar r \leftarrow \operatorname{mean}_l(r_l)$
\FOR{each prunable layer $l$}
    \STATE $\alpha_l \leftarrow \alpha^{r_l/\bar r}$
    \STATE $\mathbf{B}_l^{(0)} \leftarrow \operatorname{clip}\!\left(\alpha_l \mathbf{B}_l^{\mathrm{pre}},\, b_{\min},\, b_{\max}\right)$
\ENDFOR
\STATE \textbf{return} $\{\mathbf{B}_l^{(0)}\}$
\end{algorithmic}
\end{algorithm}

After pushing the resource toward the target budget, we apply spectral-constrained topology compensation to improve the trainability of the sparse subnet under $E^\star$.
For each prunable layer $l$, we compute the structural score $\mathbf{S}_l = |\mathbf{W}_l| \odot \mathbf{B}_l^{(0)}$ from the weight magnitude and the reallocated quantiser bit widths.
Based on $\mathbf{S}_l$, we first select the highest scoring connections under the layer budget induced by $\mathbf{B}_l^{(0)}$ and the minimum degree constraint to form an initial binary mask $\mathbf{M}_l$.
The spectral projection is implemented as a greedy local repair procedure: if the retained active matrix violates the condition number bound in Eq.~\ref{eq:sp}, selected lower score connections are iteratively replaced by unselected higher score candidates only when the replacement reduces $\kappa(\mathbf{W}_l \odot \mathbf{M}_l)$ while preserving the budget constraint.

\begin{equation}
\kappa(\mathbf{W}_l \odot \mathbf{M}_l)
=
\frac{\sigma_{\max}(\mathbf{W}_l \odot \mathbf{M}_l)}{\sigma_{\min}(\mathbf{W}_l \odot \mathbf{M}_l)+\epsilon}
\le \tau_l
\label{eq:sp}
\end{equation}

Here, $\sigma_{\max}(\cdot)$ and $\sigma_{\min}(\cdot)$ denote the largest and smallest singular values, $\epsilon$ is a small constant for numerical stability, and $\tau_l$ is set adaptively as $\tau_l = 3\,\kappa(\mathbf{W}_l^{\mathrm{pre}})$, which provides a practical layer-wise condition bound for sparse supports when each active node is constrained to have degree at least $2$.
In this check, disconnected or zero-bitwidth channels are treated as pruned paths and are excluded from the retained active matrix, so the numerical floor $\epsilon$ is not used to interpret intentionally removed channels as rank collapse.
Unlike a spectral-norm bound that only limits the largest layer gain, this check also discourages ill-conditioned active sparse supports.
The spectral criterion serves as a proxy for trainability, since it suppresses ill-conditioned sparse networks and helps preserve a stable signal and gradient path.
We then obtain the spectrally well-posed layer by projecting $\mathbf{W}_l \leftarrow \mathbf{W}_l \odot \mathbf{M}_l$ and $\mathbf{B}_l^{\star} = \mathbf{M}_l \odot \mathbf{B}_l^{(0)}$.
To close any residual resource gap, a global scalar $\lambda$ is applied uniformly to all active bit widths, and the total budget is evaluated as Eq.~\ref{eq:budget_gamma}.

\begin{equation}
E(\lambda)
=
\sum_l \sum_{i,j}
c_{l,ij} M_{l,ij}
\operatorname{clip}
\left(
\lambda B^{\star}_{l,ij},
b_{\min},
b_{\max}
\right).
\label{eq:budget_gamma}
\end{equation}

The optimal scale is obtained by minimising the absolute deviation from the target $\lambda^\star = \operatorname*{arg\,min}_{\lambda}\,
\left| E(\lambda) - E^\star \right|$.
Finally, each layer's final bit-width field is set to $\mathbf{B}_l^\star \leftarrow \operatorname{clip}(\lambda^\star \mathbf{B}_l^\star, b_{\min}, b_{\max})$.
The pruned network will then be used for fine-grained exploration near the target resource.

\subsection{Bidirectional $\beta$ scheduling}

\begin{figure*}[t]
    \centering
    \includegraphics[width=0.95\textwidth]{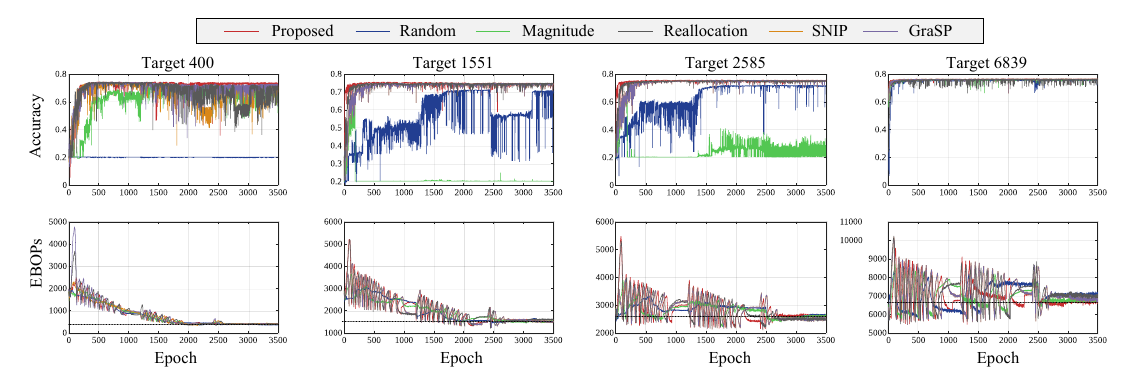}  
    \vspace{-4mm}
    \caption{Comparison of pruning trajectories under different target EBOP budgets. The upper row shows test accuracy during training, and the lower row shows the corresponding observed EBOPs. Results are compared between the proposed method and five pruning baselines, namely SNIP, Random, Reallocation, Magnitude, and GraSP, under target budgets of 400, 1551, 2585, and 6839 EBOPs.}
    \label{fig:pruning_trajectories} \vspace{-4mm}
\end{figure*}

\begin{algorithm}[h]
\caption{Bidirectional $\beta$ scheduling for near-target search}
\label{alg:bidirectional_beta}
\begin{algorithmic}[1]
\STATE \textbf{Input:} pruned model $\mathcal{N}$, target budget $E^\star$
\STATE \textbf{Output:} fine-tuned model $\mathcal{N}$ near $E^\star$
\STATE \textbf{Initialise:} $\beta \leftarrow \beta_0$
\STATE \textbf{Define:}
$\beta^{\uparrow}(\beta,E)=\min\!\left(\beta \sqrt{E/E^\star},\beta_{\mathrm{hi}}\right)$,
\STATE $\beta^{\downarrow}(\beta,E)=\max\!\left(\beta \sqrt{E/E^\star},\beta_{\mathrm{lo}}\right)$
\FOR{each epoch $t=1,2,\dots,T$}
    \STATE Train $\mathcal{N}$ for one epoch with current $\beta$
    \STATE Measure $E_t^{\mathrm{obs}}$
    \IF{$E_t^{\mathrm{obs}} > E^\star$}
        \STATE $\beta \leftarrow \beta^{\uparrow}(\beta,E_t^{\mathrm{obs}})$
    \ELSIF{$E_t^{\mathrm{obs}} < E^\star$}
        \STATE $\beta \leftarrow \beta^{\downarrow}(\beta,E_t^{\mathrm{obs}})$
    \ENDIF
    \IF{validation metric stagnates for $S$ epochs}
        \STATE $\beta \leftarrow \beta_{\mathrm{lo}} + \frac{t}{T}\left(\beta-\beta_{\mathrm{lo}}\right)$
    \ENDIF
\ENDFOR
\STATE \textbf{return} $\mathcal{N}$
\end{algorithmic}
\end{algorithm}

Bidirectional $\beta$ scheduling is then used to refine the current subnetwork around the target resource budget within a limited fine-tuning epochs, as summarised in Algorithm~\ref{alg:bidirectional_beta}. At epoch $t$, the model is trained with the current resource-pressure coefficient $\beta$, and the resulting resource usage is measured as $E_t^{\mathrm{obs}}$.
The controller then compares $E_t^{\mathrm{obs}}$ directly with the target budget $E^\star$ and updates $\beta$ through a closed-loop.

The inner loop constrains the search region around the target budget, when the observed resource $E_t^{\mathrm{obs}}$ is higher than the targeted resource budget, the update $\beta^{\uparrow}(\beta,E_t^{\mathrm{obs}})$ increases the resource pressure and pushes the optimisation back toward the target region.
Vice versa. In both cases, the adjustment strength is determined by the budget ratio $\sqrt{E_t^{\mathrm{obs}}/E^\star}$ that keeps the response stable near the target.

However, near-target optimisation can still stagnate under stable budget control. 
As $\beta$ enforces the resource constraint, many quantisers are pushed toward their lower admissible range, which restricts the trainable degrees of freedom of the current subnetwork.
Low bit widths reduce model capacity and weaken the effective straight-through gradients for quantiser readjustment, making suppressed connections difficult to reactivate.
Updates that restore bit widths or connectivity usually increase resource usage and are therefore penalised by $\beta$.
The optimisation then enters a $\beta$ deadlock, that is, the model remains near $E^\star$, but the compressed configuration becomes hard to escape and validation performance stalls.
The same mechanism also exists in monotonic resource reduction, where sustained compression pressure can lock the search into an over-suppressed subnetwork.

The outer loop monitors the validation metric over a stall window of $S$ epochs to alleviate this locking effect.
When no improvement is observed, $\beta$ is relaxed toward $\beta_{\mathrm{lo}}$ by
$\beta \leftarrow \beta_{\mathrm{lo}} + \frac{t}{T}\left(\beta-\beta_{\mathrm{lo}}\right)$.
The subsequent restart releases the compression pressure and extends the search space.
With the timing control of $t$, earlier restarts explore a broader neighbourhood, while later restarts focus on smaller local refinements.
The subsequent inner-loop updates then steer the model back toward $E^\star$.
Unlike PID-based $\beta$ control that primarily tracks a target EBOPs~\cite{laatu2025sub}, our scheduler uses the inner budget correction and the outer $\beta$ relaxation to perform annealing-style near-target search, thereby mitigating accuracy stagnation and missed better solutions caused by $\beta$ deadlock.

\section{Experimental Results}\label{results}
We conduct experiments on the 5-class JSC classification task with two datasets, OpenML~\cite{openml_lhc_jets_hlf_2020} and CERNBox~\cite{cernbox_lhc_jets_2025}, to validate the proposed algorithm. The evaluated model is a fully connected network with three hidden layers with a layer-wise configuration of 16, 64, 32, 32, 5. 
The model training and accelerator compilation are performed on a workstation equipped with an NVIDIA RTX A4000 GPU and an Intel U7 265K CPU, using CUDA 12.8, Python 3.12, da4ml~\cite{sun2025da4ml} 0.5.0 and HGQ~\cite{sun2026hgq} 0.1.6 with a TensorFlow backend.
All resource and latency results reported in this section are obtained from out-of-context place-and-route in Vivado 2025.1, targeting a Virtex Ultrascale+ FPGA (\texttt{xcvu13p-flga2577-2-e}).


\begin{figure*}[t]
    \centering
    \includegraphics[width=0.95\textwidth]{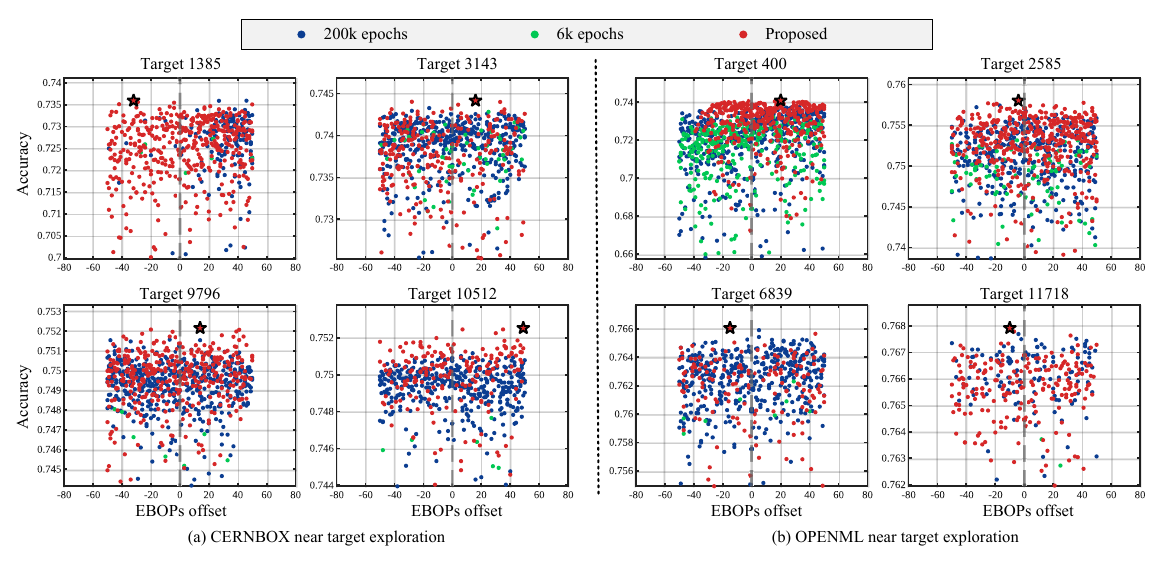}  
    \vspace{-4mm}
    \caption{Near-target exploration under different EBOP budgets on the CERNBox and OpenML datasets. Each point denotes a validation record collected during training, where the horizontal axis is the EBOPs offset from the target budget and the vertical axis is the validation accuracy. Blue, green, and red points correspond to HGQ with 200k epochs, HGQ with 6k epochs, and the proposed method, respectively.}
    \vspace{-4mm}
    \label{fig:near_target_exploration}
\end{figure*}

We take network search targeting 400 EBOPs to quantify the search efficiency.
HGQ's progressive search, as the baseline, adopts an increasing $\beta$ schedule from $5 \times 10^{-7}$ to $10^{-3}$ with the training epochs.
RQP performs one-shot quantiser pruning from a high-precision network trained for 2,500 epochs, followed by 3,500 epochs of near budget search.
Model training is performed with a seed of 1998.
The experimental results in Table~\ref{tab:search_time_400ebops} demonstrate that RQP can rapidly reach the target budget through one-shot pruning, achieving a reduced training time of $20.58\times$ over the 200,000 epochs HGQ training.
Although the progressive search with the same 6,000 epochs shows a lower training time with less control overhead, the subnets discovered by RQP through fine-grained near budget search still achieve better test accuracy and resource gap.

\begin{table}[t]
\centering
\caption{Search efficiency and model quality around 400 EBOPs.}
\label{tab:search_time_400ebops}
\begin{tabular}{lcccc}
\toprule
Impl.&
\makecell{Acc. (\%)\\@ 400 EBOPs} &
\makecell{Actual\\EBOPs} &
\makecell{Total\\Epochs} &
\makecell{Training time\\(min)} \\
\midrule
HGQ
& 72.84 {\color{red}(-0.80)} 
& 390 {\color{ForestGreen}(-10)} 
& 6,000   
& 15.27 {\color{ForestGreen}($0.84\times$)} \\

HGQ
& 72.86 {\color{red}(-0.78)} 
& 407 {\color{red}(+7)} 
& 20,000  
& 45.78 {\color{red}($2.53\times$)} \\

HGQ  
& 73.49 {\color{red}(-0.15)} 
& 396 {\color{ForestGreen}(-4)} 
& 100,000 
& 186.48 {\color{red}($10.29\times$)} \\

HGQ 
& 73.64 {\color{black}(0.00)}
& 398 {\color{ForestGreen}(-2)} 
& 200,000 
& 372.96 {\color{red}($20.58\times$)} \\

Prop. 
& 73.87 {\color{ForestGreen}(+0.23)} 
& 399 {\color{ForestGreen}(-1)} 
& 6,000   
& 18.12 {\color{black}($1\times$)} \\
\bottomrule
\end{tabular}
\vspace{-4mm}
\end{table}


The quantiser pruning adopted in RQP differs from conventional binary-mask pruning methods in that it directly operates on the per-weight quantiser bit widths, which are continuous optimisation variables beyond the reach of binary pruning, and can therefore drive the network toward the target resource budget more effectively.
For a fair comparison in this case study, we map the binary pruning masks produced by conventional pruning methods onto the quantisers and perform pruning over both the weights and the associated quantiser bit widths.
Fig.~\ref{fig:pruning_trajectories} illustrates the accuracy recovery and resource approaching trend of pruning algorithms across different target resources after applying warm-up $\beta$-scheduled search.
For higher resource budgets, the network remains relatively tolerant to the binary pruning mask.
In the lower-budget region, the subnetwork's trainability with random and magnitude pruning degrades significantly.
Across all target budgets, the subnetworks generated by RQP consistently show a stronger capability for accuracy recovery compared to those obtained by existing binary-mask pruning methods. The Reallocation variant shown in the graph is used as an ablation of the spectral constraint component, separating the contribution of the spectral constraint towards improved network trainability.


We analyse the design frontiers of the JSC classification task on two datasets to evaluate the performance of the model obtained by our method at different target resources.
The training-time, validation accuracy and the corresponding EBOPs are shown in Fig.~\ref{fig:near_target_exploration}. 
Under bidirectional $\beta$ scheduling, RQP enables denser exploration around the target resource region and, simultaneously by releasing the $\beta$ lock, discovers better frontier points within a limited training budget.

All dominating points during the training are stored as checkpoints for test accuracy evaluation and accelerator generation to compare with previous approaches.
In Fig.~\ref{fig:mainpareto}, we captured the design frontier of test accuracy and implemented LUT usage explored by the three methods on the CERNBox dataset, together with other accelerator design points~\cite{weng2025greater, cassidy2025reducedlut, andronic2025neuralut, coelho2021automatic, lou2024polylut, umuroglu2020logicnets, andronic2024neuralut, andronic2025polylut}.
It can be observed that, compared to HGQ, our method achieves a higher accuracy frontier under the similar resource usage. 
Table~\ref{tab:HGQ_prop_compare} compares the post-implementation performance and resource usage of the generated accelerators at several representative resource points.
Beyond the accuracy improvement, removing the $\beta$ locking effect also leads to different bit-width distributions in the post-pruning trained networks compared with those obtained by progressive search.
This difference is reflected in a higher achievable maximum frequency because of the reduced bit-width distribution and critical path.
Although the increase in low-bit-width computation may lead to more pipeline stages (or latency cycles when II=1) in some cases, this is mostly compensated by the higher achievable frequency for better end-to-end latency (in ns).

\begin{table}[t]
\centering
\caption{Hardware results of the designs obtained by RQP and HGQ.}
\label{tab:HGQ_prop_compare}
\begin{tabular}{lccccc}
\toprule
Imp. & Acc. (\%) & \makecell{Latency\\(cycle / ns)} & LUT & FF & \makecell{Fmax\\(MHz)} \\
\midrule
HGQ  & 73.64 & 6 (9.06 ns)   & 364    & 361    & 662.326 \\
Prop. & 73.87 & 6 (8.94 ns)   & 368    & 392    & 671.217 \\
HGQ  & 75.45 & 12 (23.02 ns) & 2,227  & 2,336  & 521.376 \\
Prop. & 75.65 & 14 (21.29 ns) & 2,189  & 2,405  & 657.462 \\
HGQ  & 76.59 & 16 (44.94 ns) & 5,942  & 5,197  & 355.999 \\
Prop. & 76.61 & 17 (44.49 ns) & 5,943  & 5,245  & 382.117 \\
HGQ  & 76.85 & 21 (42.11 ns) & 10,804 & 12,910 & 498.753 \\
Prop. & 76.87 & 21 (41.16 ns) & 11,219 & 13,196 & 510.204 \\
\bottomrule
\end{tabular}
\vspace{-5mm}
\end{table}
\section{Conclusion}\label{conclusion}
In this paper, we present RQP, a resource-oriented quantiser pruning method for HGQ-based neural networks on FPGAs.
RQP combines resource-oriented one-shot quantiser pruning with bidirectional $\beta$ scheduling to improve search efficiency near a target deployment budget.
For neural network accelerator training on the JSC classification task, RQP reaches the targeted design point up to $20.58\times$ faster than HGQ's progressive monotonic resource reduction.
In future work, we will further explore the pruning strategy, adapt the framework to multi-seed evaluation settings and expand the framework to support more architectures, including CNNs and transformers.

\bibliographystyle{unsrt}
\bibliography{references}

\end{document}